# An extremely primitive halo star

Elisabetta Caffau[1,2*], Piercarlo Bonifacio[2], Patrick François[2,3], Luca Sbordone[1,4,2], Lorenzo Monaco[5], Monique Spite[2], François Spite[2], Hans-G. Ludwig[1,2], Roger Cayrel[2], Simone Zaggia[6], François Hammer[2], Sofia Randich[7], Paolo Molaro[8], Vanessa Hill[9]

[1]*Zentrum für Astronomie der Universität Heidelberg, Landessternwarte, Königstuhl 12, 69117 Heidelberg, Germany*
[2]*GEPI, Observatoire de Paris, CNRS, Université Paris Diderot, Place Jules Janssen, 92190 Meudon, France*
[3]*UPJV, Université de Picardie Jules Verne, 33 Rue St Leu, F-80080 Amiens*
[4]*Max-Planck Institut für Astrophysik, Karl-Schwarzschild-Str. 1, 85741 Garching, Germany*
[5]*European Southern Observatory, Casilla 19001, Santiago, Chile*
[6]*Istituto Nazionale di Astrofisica, Osservatorio Astronomico di Padova Vicolo dell'Osservatorio 5, 35122 Padova, Italy*
[7]*Istituto Nazionale di Astrofisica, Osservatorio Astrofisico di Arcetri, Largo E. Fermi 5, 50125 Firenze, Italy*
[8]*Istituto Nazionale di Astrofisica, Osservatorio Astronomico di Trieste, Via Tiepolo 11, 34143 Trieste, Italy*
[9]*Université de Nice-Sophia Antipolis, Observatoire de la Côte d'Azur, CNRS, Laboratoire Cassiopée, bd de l'Observatoire, 06300 Nice, France*

**The early Universe had a chemical composition consisting of hydrogen, helium and traces of lithium[1], almost all other elements were created in stars and supernovae. The mass fraction, Z, of elements more massive than helium, is called "metallicity". A number of very metal-poor stars have been found[2,3], some of which, while having a low iron abundance, are rich in carbon, nitrogen and oxygen[4,5,6]. For theoretical reasons[7,8] and because of an observed absence of stars with metallicities lower than $Z=1.5\times10^{-5}$, it has been suggested that low mass stars ($M<0.8 M_\odot$, the ones that survive to the present day) cannot form until the interstellar medium has been enriched above a critical value, estimated to lie in the range $1.5\times10^{-8} \le Z \le 1.5\times10^{-6}$[8], although competing**

---
* Gliese Fellow

**theories claiming the contrary do exist[9]. Here we report the chemical composition of a star with a very low $Z \leq 6.9 \times 10^{-7}$ ($4.5 \times 10^{-5}$ of that of the Sun[10]) and a chemical pattern typical of classical extremely metal poor stars[2,3], meaning without the enrichment of carbon, nitrogen and oxygen. This shows that low-mass stars can be formed at very low metallicity. Lithium is not detected, suggesting a low metallicity extension of the previously observed trend in lithium depletion[11]. Lithium depletion implies that the stellar material must have experienced temperatures above two million K in its history, which points to rather particular formation condition or internal mixing process, for low Z stars.**

The star SDSS J102915+172927 (RA = 10h 29m 15.15s and declination $\delta = +17° 29' 28''$ at equinox 2000, g magnitude 16.92, (g-z)=0.59, (g-z)$_0$=0.53) object of this letter, has been observed with the X-Shooter[12] and UVES[13] spectrographs at VLT, facilities of the European Southern Observatory in Chile. Theoretical model atmospheres and spectrum synthesis techniques have been used to derive the chemical abundances provided in Table 1. The chemical signatures are consistent with metal production by ordinary core-collapse supernovae[14]. The derived abundances coupled with the upper limits on carbon and nitrogen imply $Z \leq 6.9 \times 10^{-7}$. This number takes into account the typical "excess" of the alpha-element oxygen, [O/Fe]=+0.6. Our analysis has been performed assuming local thermodynamic equilibrium (LTE), further work is necessary to assess the role of departures from LTE, especially for molecules. The estimate of NLTE effects on magnesium[15] is about +0.4 dex, which translates in a change of +0.2 ×$10^{-7}$ in Z.

It has been suggested that the primary discriminant between the formation of only massive stars (as in population III) and of both massive and low-mass stars (as in populations II and I) are the abundances of carbon and oxygen[7]. This is because these elements can provide efficient cooling of the proto-stellar clouds through the fine structure lines of ionised carbon and neutral oxygen. A suitable combination of the carbon and oxygen abundances is called the transition discriminant[16] [D=log10($10^{[C/H]}$+0.3×$10^{[O/H]}$)], and low mass star formation is believed to occur only if D ≥-3.5. From the abundances in Table 1 and the assumption [O/Fe]=+0.6, we have for SDSS J102915+172927, D≤-4.2, which places it in the ``forbidden zone'' of the theory. If, instead of taking the upper limit on the carbon abundance, we assume that the carbon abundance, derived from the 3D analysis, scales with the iron abundance, as found in other metal-poor stars[3], we have D≤-4.4. Our measurement cannot rule out the above mentioned theoretical scenario[7,16], but it strongly supports the idea that, at least in some cases, low mass stars can also form at lower carbon and oxygen abundances than the current estimates for the critical values.

The complete absence of the lithium resonance doublet at 670.7 nm is remarkable. In fact most of the "warm" metal-poor dwarf stars display a constant abundance of lithium, the so-called Spite plateau[11,17]. From the signal-to-noise ratio in the UVES spectrum of SDSS J102915+172927, we derive an upper-limit for the Li abundance A(Li)<1.1 (at 5 σ). In Fig. 2 we show the Spite plateau as a function of the carbon abundance, as well as a function of the iron abundance, which we use in turn as a proxy of the metallicity Z. The sample of stars is composed of stars with a normal carbon abundance[3,18,19,20] and the carbon-rich, iron-poor subgiant HE 1327-2326[4]. The picture emerging from the two panels shows the same morphology, with the exception of star HE 1327-2326, with [Fe/H] lower than all the others, but [C/H] comparable to many other stars in the sample. It is noteworthy that the only two stars with [Fe/H]<-4.5 have no detectable lithium.

The most straightforward interpretation of the Spite plateau is that the lithium observed in the plateau stars is the lithium produced in the big bang[17]. The theoretical primordial Li abundance[1] is a factor of 2 to 3 larger than the value observed on the Spite plateau. A number of explanations of this discrepancy have been proposed, which range from stellar phenomena, such as atomic diffusion[21], to new physics leading to a different big bang nucleosynthesis[22]. Our upper limit implies that the Li abundance of SDSS J102915+172927 is far below the value of the Spite plateau. At extremely low metallicities, the Spite plateau displays a "meltdown" [11], i.e. an increased scatter and a lower mean Li abundance. This meltdown is clearly seen in the two components of the extremely metal-poor binary system CS 22876-32 that show a different Li content[19]. The primary is on the Spite plateau, while the secondary is below at A(Li)=1.8. The reasons for this meltdown are not understood. It has been suggested[11] that a Li depletion mechanism, whose efficiency is metallicity and temperature dependent, could explain the observations. If this were the case, the Li abundance in SDSS J102915+172927 would result from an efficient Li depletion due to a combination of extremely low metallicity and relatively low temperature. For completeness we mention that there is a small number of known stars which have a metallicity and effective temperature similar to that of other stars on the Spite plateau, but where the Li doublet is not detected. The fact that such stars are found for different values of [Fe/H] and [C/H] suggests that Li-depletion is independent of either. It has been suggested that Li-depleted stars could have a common origin with blue stragglers[23], an interpretation that has been reinforced by the discovery that these stars are also depleted in beryllium[24].

Stars similar to SDSS J102915+172927 are probably not so rare. Only 30% of the whole SDSS survey area was accessible to our VLT observations. We identified 2899 potentially extreme stars with metallicity less than $Z \leq 1.1\times10^{-5}$ in Data Release 7[25]. Among those observable with the VLT we performed a subjective selection of the most promising candidates of which we observed six in our X-Shooter programme resulting in one detection. Depending on the subjective bias we attribute to the last selection step, we expect 5 to 50 stars of similar or even lower metallicity than SDSS J102915+172927 to be found among the candidates accessible from the VLT, and many more in the whole SDSS sample.

## References

1. Iocco, F., Mangano, G., Miele, G., Pisanti, O., Serpico, P. D. Primordial nucleosynthesis: From precision cosmology to fundamental physics. *Physics Reports*, 472, 1-76 (2009)

2. Cayrel, R., Depagne, E., Spite, M., Hill, V., Spite, F., François, P., Plez, B., Beers, T., Primas, F., Andersen, J., Barbuy, B., Bonifacio, P., Molaro, P., Nordström, B. First stars V - Abundance patterns from C to Zn and supernova yields in the early Galaxy *Astron. Astrophys.*, 416, 1117-1138 (2004)

3. Bonifacio, P., Spite, M., Cayrel, R., Hill, V., Spite, F., François, P., Plez, B., Ludwig, H.-G., Caffau, E., Molaro, P., Depagne, E., Andersen, J., Barbuy, B., Beers, T. C., Nordström, B., Primas, F. First stars XII. Abundances in extremely metal-poor turnoff stars, and comparison with the giants. *Astron. Astrophys.*, 501, 519-530 (2009)

4. Frebel, A., Collet, R., Eriksson, K., Christlieb, N., Aoki, W. HE 1327-2326, an Unevolved Star with [Fe/H]<-5.0. II. New 3D-1D Corrected Abundances from a Very

Large Telescope UVES Spectrum. *Astrophys. J.*, 684, 588-602 (2008)

5. Christlieb, N., Bessell, M. S., Beers, T. C., Gustafsson, B., Korn, A., Barklem, P. S., Karlsson, T., Mizuno-Wiedner, M., Rossi, S. A stellar relic from the early Milky Way *Nature*, 419, 904-906 (2002)

6. Norris, J. E., Christlieb, N., Korn, A. J., Eriksson, K., Bessell, M. S., Beers, T. C., Wisotzki, L., Reimers, D. HE 0557-4840: Ultra-Metal-Poor and Carbon-Rich *Astrophys. J.*, 670, 774-788 (2007)

7. Bromm, V., Loeb, A. The formation of the first low-mass stars from gas with low carbon and oxygen abundances *Nature*, 425, 812-814 (2003)

8. Schneider, R., Ferrara, A., Salvaterra, R., Omukai, K., Bromm, V. Low-mass relics of early star formation, *Nature*, 422, 869-871 (2003)

9. Nakamura, F., Umemura, M. On the Initial Mass Function of Population III Stars *Astrophys. J.*, 548, 19-32 (2001)

10. Caffau, E., Ludwig, H.-G., Steffen, M., Freytag, B., Bonifacio, P. Solar Chemical Abundances Determined with a CO5BOLD 3D Model Atmosphere. *Solar Physics*, 268, 255-269 (2011)

11. Sbordone, L., Bonifacio, P., Caffau, E., Ludwig, H.-G., Behara, N. T., González Hernández, J. I., Steffen, M., Cayrel, R., Freytag, B., van't Veer, C., Molaro, P., Plez, B., Sivarani, T., Spite, M., Spite, F., Beers, T. C., Christlieb, N., François, P., Hill, V. The metal-poor end of the Spite plateau. I. Stellar parameters, metallicities, and lithium abundances. *Astron. Astrophys.*, 522, A26-(2010)

12. D'Odorico, S., Dekker, H., Mazzoleni, R., Vernet, J., Guinouard, I., Groot, P., Hammer, F., Rasmussen, P. K., Kaper, L., Navarro, R., Pallavicini, R., Peroux, C., Zerbi, F. M. X-shooter UV- to K-band intermediate-resolution high-efficiency spectrograph for the VLT: status report at the final design review. *Society of Photo-Optical Instrumentation Engineers (SPIE) Conference Series*, 6269, (2006)

13. Dekker, H., D'Odorico, S., Kaufer, A., Delabre, B., Kotzlowski, H. Design, construction, and performance of UVES, the echelle spectrograph for the UT2 Kueyen Telescope at the ESO Paranal. *Observatory Society of Photo-Optical Instrumentation Engineers (SPIE) Conference Series*, 4008, 534-545 (2000)

14. Chieffi, A., Limongi, M. The Explosive Yields Produced by the First Generation of Core Collapse Supernovae and the Chemical Composition of Extremely Metal Poor Stars. *Astrophys. J.*, 577, 281-294 (2002)

15. Andrievsky, S. M., Spite, M., Korotin, S. A., Spite, F., Bonifacio, P., Cayrel, R., François, P., Hill, V. Non-LTE abundances of Mg and K in extremely metal-poor stars and the evolution of [O/Mg], [Na/Mg], [Al/Mg], and [K/Mg] in the Milky Way *Astron. Astrophys.*, 509, A88-(2010)

16. Frebel, A., Johnson, J. L., Bromm, V. Probing the formation of the first low-mass stars with stellar archaeology, *Mon. Not. R. Astron. Soc.*, 380, L40-L44 (2007)

17. Spite, M., Spite, F. Lithium abundance at the formation of the Galaxy. *Nature*, 297, 483-485 (1982)

18. Bonifacio, P., Molaro, P., Sivarani, T., Cayrel, R., Spite, M., Spite, F., Plez, B., Andersen, J., Barbuy, B., Beers, T. C., Depagne, E., Hill, V., François, P., Nordström,

B., Primas, F. First stars VII - Lithium in extremely metal poor dwarfs. *Astron. Astrophys.*, 462, 851-864 (2007)

19. González Hernández, J. I., Bonifacio, P., Ludwig, H.-G., Caffau, E., Spite, M., Spite, F., Cayrel, R., Molaro, P., Hill, V., François, P., Plez, B., Beers, T. C., Sivarani, T., Andersen, J., Barbuy, B., Depagne, E., Nordström, B., Primas, F. First stars XI. Chemical composition of the extremely metal-poor dwarfs in the binary CS 22876-032. *Astron. Astrophys.*, 480, 233-246 (2008)

20. Norris, J. E., Ryan, S. G., Beers, T. C., Deliyannis, C. P. Extremely Metal-Poor Stars. III. The Li-depleted Main-Sequence Turnoff Dwarfs. *Astrophys. J.*, 485, 370-379(1997)

21. Richard, O., Michaud, G., Richer, J. Implications of WMAP Observations on Li Abundance and Stellar Evolution Models Astrophys. J., 619, 538-548 (2005)

22. Jedamzik, K., Pospelov, M. Big Bang nucleosynthesis and particle dark matter *New Journal of Physics*, 11, 105028-(2009)

23. Ryan, S. G., Gregory, S. G., Kolb, U., Beers, T. C., Kajino, T. Rapid Rotation of Ultra-Li-depleted Halo Stars and Their Association with Blue Stragglers. *Astrophys. J.*, 571, 501-511 (2002)

24. Boesgaard, A. M. Beryllium in Ultra-Lithium-Deficient Halo Stars: The Blue Straggler Connection *Astrophys. J.*, 667, 1196-1205 (2007)

25. Abazajian, K. N., et al. The Seventh Data Release of the Sloan Digital Sky Survey *Astrophys. J. Suppl.*, 182, 543-558 (2009)

26. Ludwig, H.-G., Bonifacio, P., Caffau, E., Behara, N. T., Gonzalez-Hernandez, J. I., Sbordone, L. Extremely metal-poor stars from the SDSS. *Physica Scripta,* Volume T, 133, 014037-(2008)

27. Kurucz, R. L. ATLAS12, SYNTHE, ATLAS9, WIDTH9, et cetera. *Memorie della Società Astronomica Italiana Supplementi*, 8, 14-24(2005)

28. Ludwig, H.-G., Caffau, E., Steffen, M., Freytag, B., Bonifacio, P., Kuĉinskas, A. The CIFIST 3D model atmosphere grid. *Memorie della Società Astronomica Italiana*, 80, 711-714(2009)

29. François, P., Depagne, E., Hill, V., Spite, M., Spite, F., Plez, B., Beers, T. C., Andersen, J., James, G., Barbuy, B., Cayrel, R., Bonifacio, P., Molaro, P., Nordström, B., Primas, F. First stars. VIII. Enrichment of the neutron-capture elements in the early Galaxy. *Astron. Astrophys.*, 476, 935-950 (2007)

Table 1 The abundances of SDSS J102915+172927 as derived from our UVES spectra. The last column provides the adopted Solar abundances on the scale $A(X)=\log_{10}(X/H)+12$. The atmospheric parameters adopted are $T_{\mathrm{eff}}$ = 5811 K, log g = 4.0 [cgs], and microturbulent velocity 1.5 km/s. The effective temperature was derived from the $(g-z)_0$ colour-$T_{\mathrm{eff}}$ calibration[26]. The combination of photometric and reddening uncertainties cause an uncertainty on $T_{\mathrm{eff}}$ of 100 K, the corresponding uncertainty on [Fe/H] (where $[A/B] = \log(N_A/N_B)_* - \log(N_A/N_B)_\odot$, and $\odot$ refers to the Sun) is 0.06 dex. We cross-checked the effective temperatures with a fit of the Halpha wings which provided the same effective temperature

within 10 K. The surface gravity has been fixed from the Balmer jump, as measured by the (u-g) colour. Other gravity indicators, such as the calcium ionisation equilibrium and the wings of higher order Balmer lines, are consistent with this choice. The uncertainty on the surface gravity is about 0.2 dex, and we can robustly exclude a surface gravity log g = 3.0 or lower, thus excluding that the star could be on the Horizontal Branch. We computed synthetic spectra with the SYNTHE code[27] and a 1D model atmosphere computed with the ATLAS 9 code[27]. These synthetic spectra were used to perform line-profile fitting for all the measurable features. The 3D corrections were computed using a 3D model atmosphere from the CIFIST grid[28] with Teff = 5850 K, log g = 4.0, and metallicity $2.7 \times 10^{-5}$. We were able to measure the abundances of only some alpha elements (Mg, Ca, Si, Ti) and two iron peak elements (Fe and Ni). The derived iron abundance is [Fe/H]=-4.99 (see Table 1, the 3D-corrected abundances should be used, the 1D abundances are given for reference only). The alpha-elements are slightly enhanced relative to iron, [Mg/Fe]=+0.4. The Sr II line at 407.8 nm is not convincingly detected, giving an upper limit [Sr/Fe]≤-0.21 which is compatible with the general pattern of low [Sr/Fe] found in extremely metal-poor stars[29]. The upper limits on carbon and nitrogen are derived by fitting the G-band around 430 nm and the NH-band around 336 nm, respectively. Unfortunately no measurement of oxygen is possible in the available spectral range neither from atomic nor from molecular lines, but there is no reason to suspect that a star not enhanced in Mg, C, and N, should be over-abundant in oxygen.

**Table 1 Abundances of SDSS J102915+172927 from high resolution spectra.**

| Element | A(X)3D | [X/H] 3D | [X/Fe]3D | [X/H] 1D | Number of lines | A(X)⊙ |
|---|---|---|---|---|---|---|
| C | ≤4.2 | ≤-4.3 | ≤+0.7 | ≤-3.8 | G-band | 8.50 |
| N | ≤3.1 | ≤-4.8 | ≤+0.2 | ≤-4.1 | NH-band | 7.86 |
| Mg I | 2.95 | -4.59±0.10 | +0.40 | -4.68±0.08 | 4 | 7.54 |
| Si I | 3.25 | -4.27±0.10 | +0.72 | -4.27±0.10 | 1 | 7.52 |
| Ca I | 1.53 | -4.80±0.10 | +0.19 | -4.72±0.10 | 1 | 6.33 |
| Ca II | 1.48 | -4.85±0.11 | +0.14 | -4.71±0.11 | 3 | 6.33 |
| Ti II | 0.14 | -4.76±0.11 | +0.23 | -4.75±0.11 | 6 | 4.90 |
| Fe I | 2.53 | -4.99±0.12 | +0.00 | -4.73±0.13 | 44 | 7.52 |
| Ni I | 1.35 | -4.88±0.11 | +0.11 | -4.55±0.14 | 10 | 6.23 |
| Sr II | ≤-2.28 | ≤-5.2 | ≤-0.21 | ≤-5.1 | 1 | 2.92 |

**Figure 1. Observed spectrum of SDSS J102915+172927.** The spectral region of the Ca II H and K lines is shown, compared to synthetic spectra computed with a global metallicity of $Z=1.1\times10^{-6}$ and solar proportions, except for alpha elements that are enhanced by 0.4 dex over iron. At the top the X-Shooter spectrum (shifted vertically by one unit for clarity) and at the bottom the UVES spectrum. The absorption due to interstellar gas is clearly detectable both in K and H Ca II lines. The measured radial velocity is -34.5±1.0 km/s .We computed a Galactic orbit from the kinematic data and the distance of 1.27±0.15 kpc, estimated from the photometry, confirming that the star belongs to the Milky Way Halo.

**Figure 2: Lithium abundance of SDSS J102915+172927 compared to that of other metal-poor stars.** The Spite plateau is shown as a function of carbon abundance, [C/H] (lower panel), and [Fe/H] (upper panel), which we use in turn as a proxy of the metallicity Z. The upper limits for SDSS J102915+172927 are from the present work. The black circles are from Ref. [3,18], the upper limit for HE 1327-2326 is from Ref. [4]. The Li measurements for the binary system CS 22876-32 are from Ref. [19] the upper limits for the three well known Li-depleted dwarfs, G122-69, G139-8 and G186-26 are from Ref. [20]. For these three latter stars, as well as for CS 22876-32 we have no measurement of the C abundances, we therefore assumed that C scales with iron as in the rest of the sample[3]. The precise placement of these stars along the abscissa in this diagram is of no consequence for the present discussion.

**Acknowledgements**

The spectra were secured through X-Shooter Guaranteed Time Observations and ESO Director's Discretionary Time at the ESO VLT Kueyen 8.2 m telescope.

**Author Contribution**

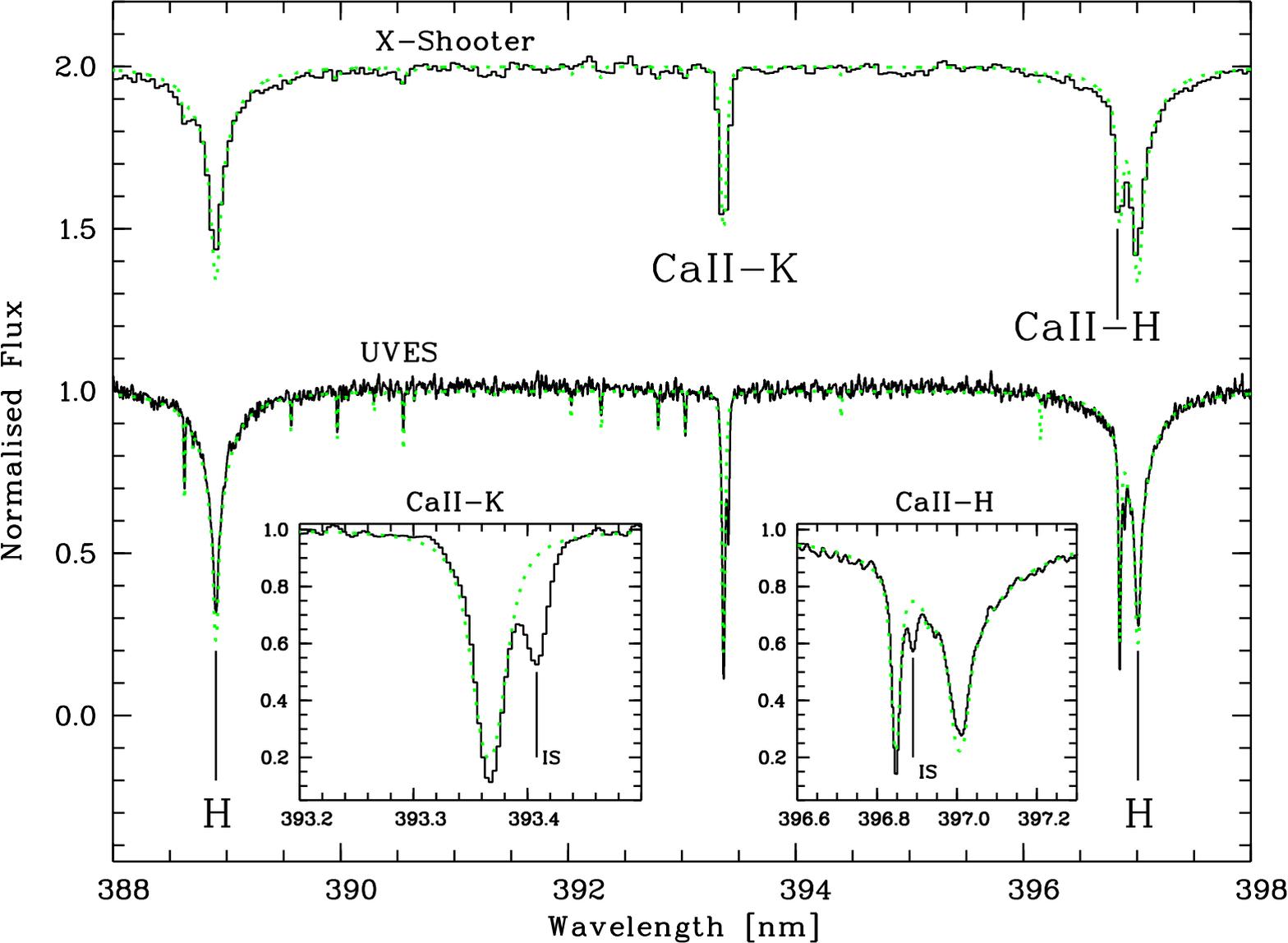

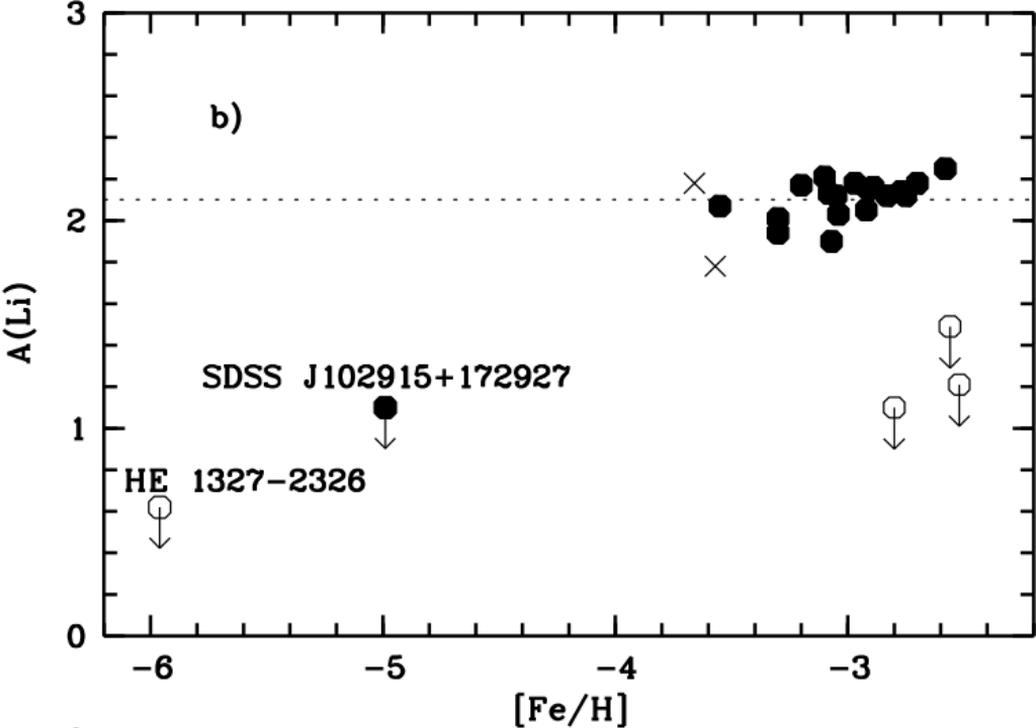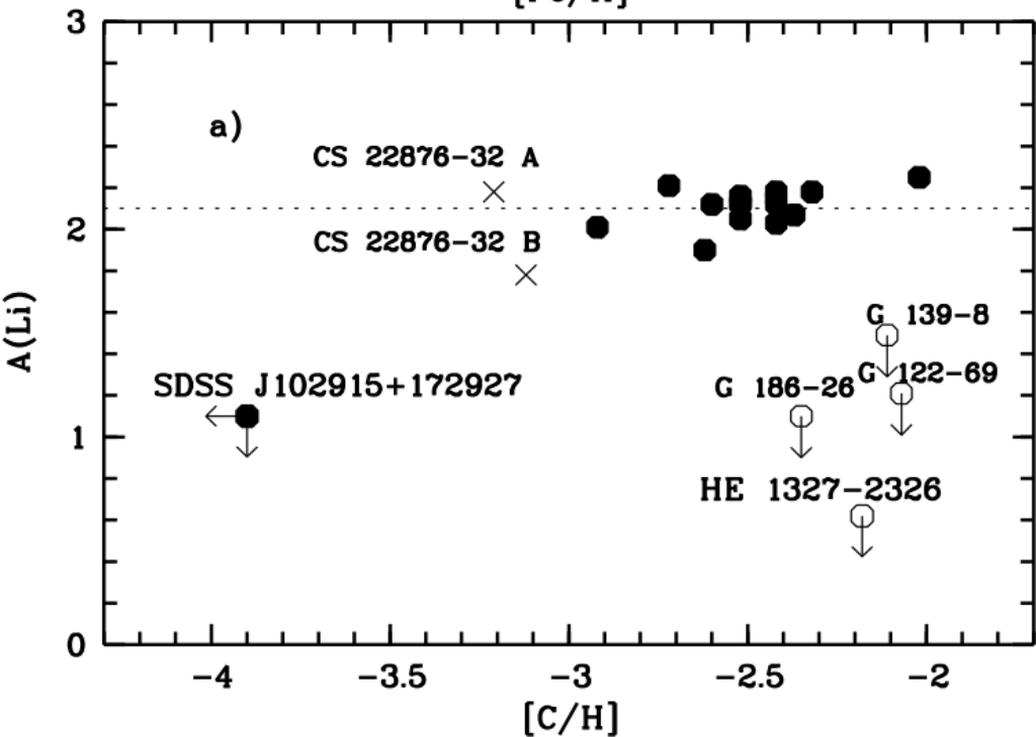